\documentclass[11pt,a4paper]{article}
\usepackage[hyperref]{acl2020}
\usepackage{times}
\usepackage{latexsym}
\usepackage{arydshln}

\usepackage{amssymb}
\usepackage{multirow}
\usepackage{amsfonts}       
\usepackage{nicefrac} 
\usepackage{latexsym}
\usepackage{multirow}
\usepackage{booktabs}
\usepackage{amsmath}
\usepackage{booktabs}
\usepackage{subfig}
\usepackage{tikz}
\usepackage{pgfplots}
\usepackage{caption}
\definecolor{mypink}{rgb}{0.9, 0.1, 0.94}
\definecolor{myyellow}{rgb}{0.98, 0.94, 0.75}
\definecolor{myblue}{rgb}{0.1, 0.9, 0.9}

\usepackage{microtype}

\aclfinalcopy 
\def\aclpaperid{3142} 

\title{CitationIE: Leveraging the Citation Graph for Scientific Information Extraction}

\author{Vijay Viswanathan, Graham Neubig, Pengfei Liu \\
  Language Technologies Institute, Carnegie Mellon University \\
  \texttt{\{vijayv, gneubig, pfliu3\}@cs.cmu.edu} \\}

\date{}

\begin{document}
\maketitle
\begin{abstract}
Automatically extracting key information from scientific documents has the potential to help scientists work more efficiently and accelerate the pace of scientific progress.
Prior work has considered extracting document-level entity clusters and relations end-to-end from raw scientific text, which can improve literature search and help identify methods and materials for a given problem.
Despite the importance of this task, most existing works on scientific information extraction (SciIE) consider extraction solely based on the content of an individual paper, without considering the paper's place in the broader literature.
In contrast to prior work, we augment our text representations by leveraging a complementary source of document context: the citation graph of referential links between citing and cited papers.
On a test set of English-language scientific documents, we show that simple ways of utilizing the structure and content of the citation graph can each lead to significant gains in different scientific information extraction tasks. When these tasks are combined, we observe a sizable improvement in end-to-end information extraction over the state-of-the-art, suggesting the potential for future work along this direction. We release software tools to facilitate citation-aware SciIE development.\footnote{\url{https://github.com/viswavi/ScigraphIE}}
\end{abstract}


\section{Introduction}

\begin{figure}[t]
  \centering
    \includegraphics[width=0.5\textwidth]{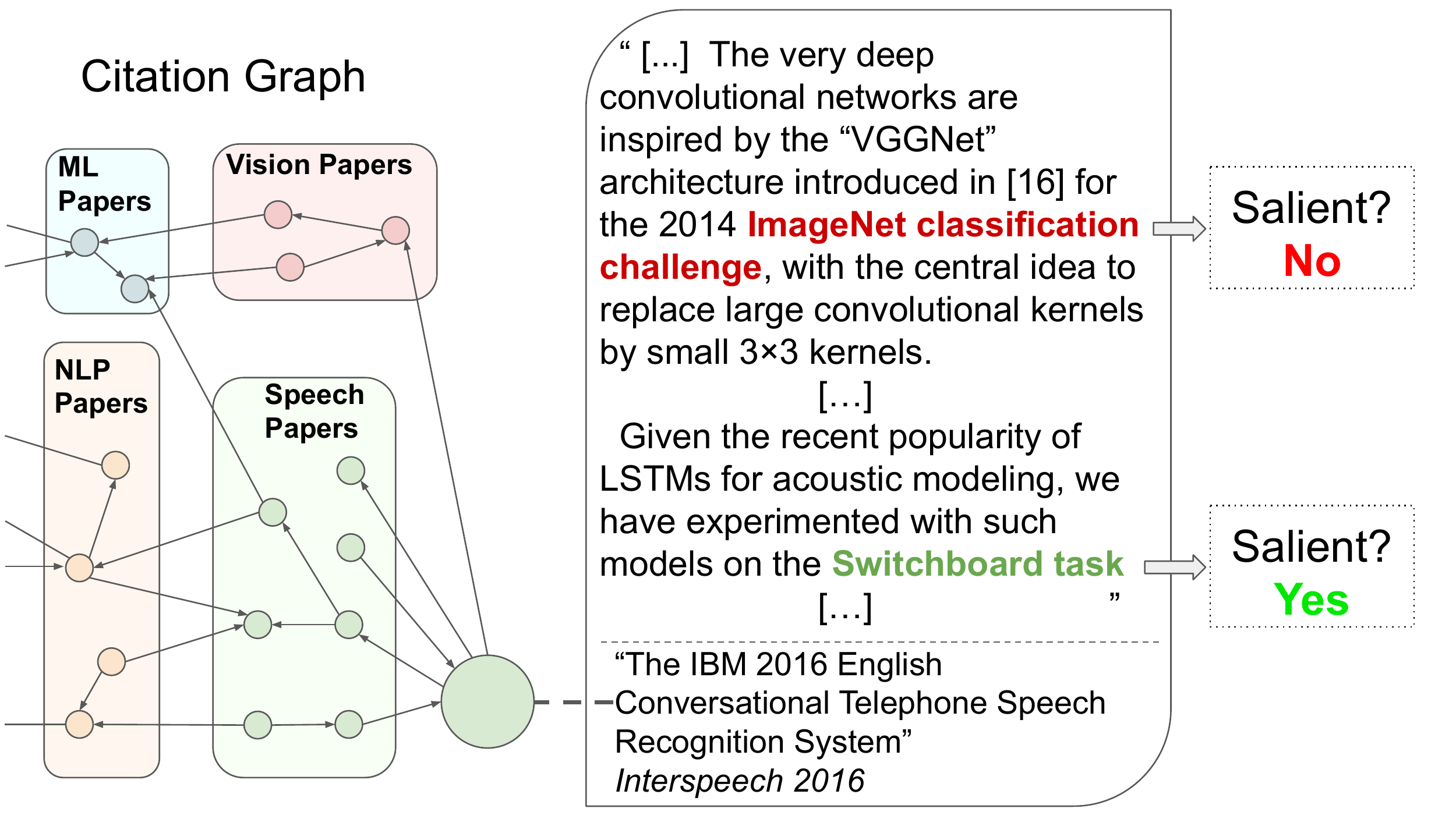}
  \caption{Example of using the citation graph to improve the task of salient entity classification \cite{jain-etal-2020-scirex}. In this task, each entity in the document is classified as salient or not, where a \emph{salient} entity is defined as being relevant to its paper's main ideas.}
\label{case_study_example}
\end{figure}


The rapid expansion in published scientific knowledge has enormous potential for good, if it can only be harnessed correctly. For example, during the first five months of the global COVID-19 pandemic, at least 11000 papers were published online about the novel disease \citep{hallenbeck-2020}, with each representing a potential faster end to a global pandemic and saved lives. Despite the value of this quantity of focused research, it is infeasible for the scientific community to read this many papers in a time-critical situation, and make accurate judgements to help separate signal from the noise.

To this end, how can machines help researchers quickly identify relevant papers? One step in this direction is to automatically extract and organize scientific information (e.g. important concepts and their relations) from a collection of research articles, which could help researchers identify new methods or materials for a given task. Scientific information extraction (SciIE) \cite{gupta-manning-2011-analyzing,yogatama-etal-2011-predicting},
which aims to extract structured information from scientific articles, has seen growing interest recently, as reflected in the rapid evolution of systems and datasets \cite{luan-etal-2018-multi,gabor-etal-2018-semeval,jain-etal-2020-scirex}.

Existing works on SciIE revolve around extraction solely based on the content of different parts of an individual paper, such as the abstract or conclusion \cite{augenstein-etal-2017-semeval,luan-etal-2019-general}.
However, scientific papers do not exist in a vacuum --- they are part of a larger ecosystem of papers, related to each other through different conceptual relations.
In this paper, we claim a better understanding of a research article relies not only on its content but also on its relations with associated works, using both the content of related papers and the paper's position in the larger citation network.


We use a concrete example to motivate how information from the citation graph helps with SciIE, considering the task of identifying key entities in a long document (known as ``salient entity classification'') in Figure~\ref{case_study_example}.

In this example, we see a paper describing a speech recognition system \cite{Saon2016TheI2}. Focusing on two specific entities in the paper (``ImageNet classification challenge" and ``Switchboard task"), we are tasked with classifying whether each is critical to the paper. This task requires reasoning about each entity in relation to the central topic of the paper, which is a daunting task for NLP considering that this paper contains over 3000 words across 11 sections.
An existing state-of-the-art model \cite{jain-etal-2020-scirex} mistakenly predicts the \emph{non-salient} entity ``ImageNet classification challenge'' as \emph{salient} due to the limited contextual information.
However, this problem is more approachable when informed of the structure of the citation graph that conveys how this paper correlates with other research works. Examining this example paper's position in the surrounding citation network suggests it is concerned with speech processing, which makes it unlikely that ``ImageNet" is salient.\footnote{Our proposed method actually makes correct predictions on both these samples, where the baseline model fails on both.}

The clear goal of incorporating inter-article information, however, is hindered by a resource challenge: existing SciIE datasets that annotate papers with rich entity and relation information fail to include their references in a fine-grained, machine-readable way. 
To overcome this difficulty, we build on top of an existing SciIE dataset and align it with a source of citation graph information, which finally allows us to explore citation-aware SciIE.

Architecturally, we adopt the neural multi-task model introduced by \citet{jain-etal-2020-scirex}, and establish a proof of concept by comparing simple ways of incorporating the network structure and textual content of the citation graph into this model.
Experimentally, we rigorously evaluate our methods, which we call \emph{CitationIE}, on three tasks: mention identification, salient entity classification, and document-level relation extraction. We find that leveraging citation graph information provides significant improvements in the latter two tasks, including a \textbf{10 point improvement} on F1 score for relation extraction. This leads to a sizable increase in the performance of the end-to-end CitationIE system relative to the current state-of-the-art, \citet{jain-etal-2020-scirex}.
We offer qualitative analysis of why our methods may work in  \S\ref{Sec:analysis}.

\section{Document-level Scientific IE}
\subsection{Task Definition}
\label{sec:task_definition}
We consider the task of extracting document-level relations from scientific texts.

Most work on scientific information extraction has used annotated datasets of
scientific abstracts, such as those provided for SemEval 2017 and SemEval 2018 shared tasks \cite{augenstein-etal-2017-semeval,gabor-etal-2018-semeval}, the SciERC dataset \cite{luan-etal-2018-multi}, and the BioCreative V Chemical Disease Relation dataset \cite{biocreative}. 

We focus on the task of open-domain document-level relation extraction from long, full-text documents. This is in contrast to the above methods that only use paper abstracts. Our setting is also different from works that consider a fixed set of candidate relations \citep{hou-etal-2019-identification, kardas-etal-2020-axcell} or those that only consider IE tasks other than relation extraction, such as entity recognition \citep{craft-corpus}.

We base our task definition and baseline models on the recently released SciREX dataset \cite{jain-etal-2020-scirex}, which contains 438 annotated papers,\footnote{The dataset contains 306 documents for training, 66 for validation, and 66 for testing.} all related to machine learning research.

Each document consists of sections $D = \{S_1, \ldots, S_N\}$, where each section contains a sequence of words $S_i = \{w_{i,1}, \ldots, w_{i, N_i}\}$.
Each document comes with annotations of entities, coreference clusters, cluster-level saliency labels, and 4-ary document-level relations.
We break down the end-to-end information extraction process as a sequence of these four related tasks, with each task taking the output of the preceding tasks as input.

\paragraph{Mention Identification}
For each span of text within a section, this task aims to recognize if the span describes a \texttt{Task}, \texttt{Dataset}, \texttt{Method}, or \texttt{Metric} entity, if any.

\paragraph{Coreference}
This task requires clustering all entity mentions in a document such that, in each cluster, every mention refers to the same entity \cite{Varkel2020PretrainingMR}. The SciREX dataset includes coreference annotations for each \texttt{Task}, \texttt{Dataset}, \texttt{Method}, and \texttt{Metric} mention. 

\paragraph{Salient Entity Classification}
Given a cluster of mentions corresponding to the same entity, the model must predict whether the entity is key to the work described in a paper. We follow the definition from the SciREX dataset \citep{jain-etal-2020-scirex}, where an entity in a paper is deemed salient if it plays a role in the paper's evaluation.

\paragraph{Relation Extraction}
The ultimate task in our IE  pipeline is relation extraction. We consider relations as 4-ary tuples of typed entities 
$(E_{\texttt{Task}}, E_{\texttt{Dataset}}, E_{\texttt{Method}}, E_{\texttt{Metric}})$, which are required to be salient entities. Given a set of candidate relations, we must determine which relations are contained in the main result of the paper. 

\begin{figure*}[t]
    \centering
    \hspace{-0.6cm}
    \includegraphics[width=0.85\linewidth]{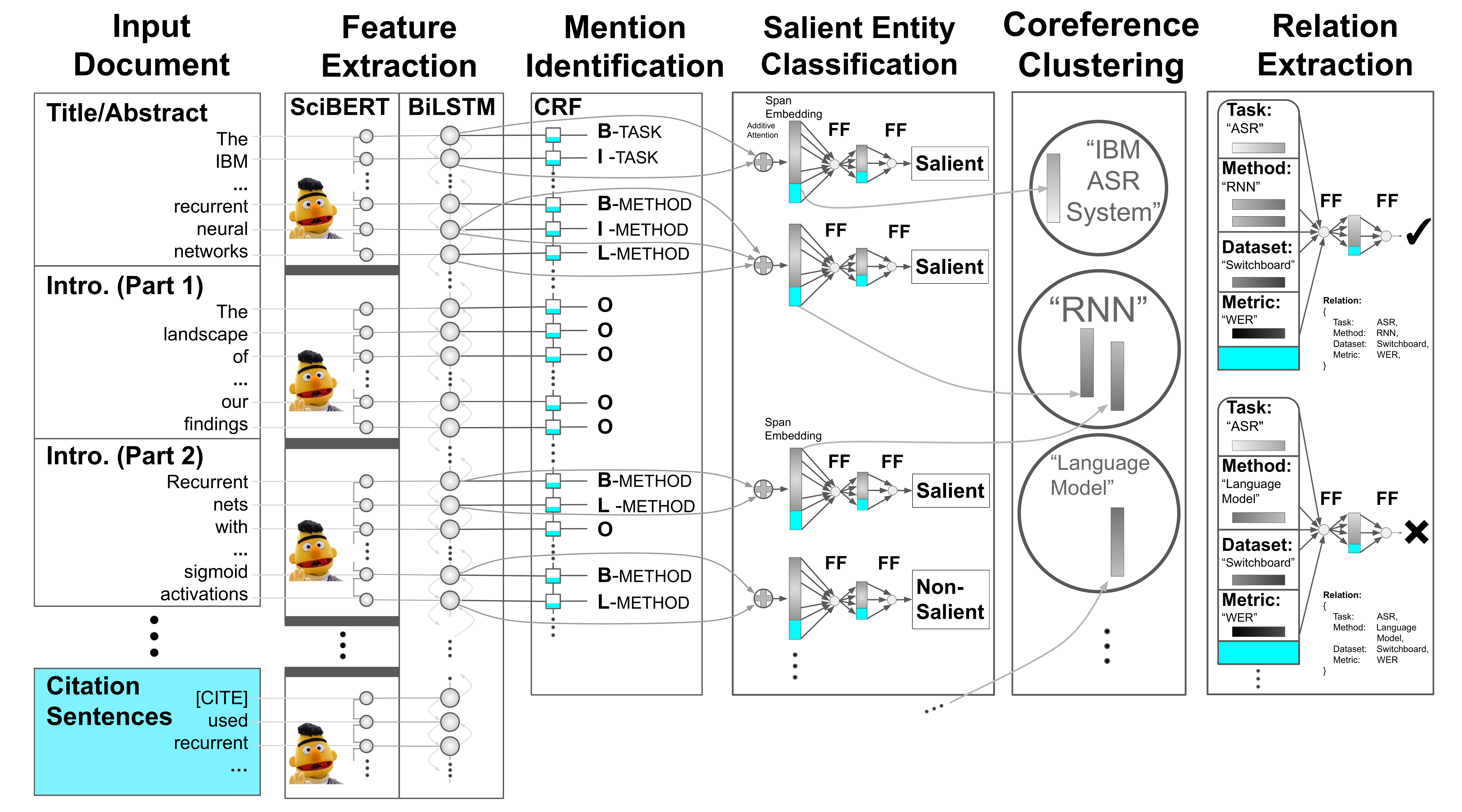}
    \caption{Architecture of the model we use for neural information extraction. Light blue blocks indicate places where we can incorporate information from the citation graph for the citation-aware CitationIE architecture.
    }
    \label{model_architecture_diagram}
\end{figure*}

\subsection{Baseline Model}
\label{section:baseline_model}
We base our work on top of the model of \citet{jain-etal-2020-scirex}, which was introduced as a strong baseline accompanying the SciREX dataset. We refer the reader to their paper for full architectural details, and briefly summarize their model here.

This multi-task model performs three of our tasks (mention identification, saliency classification, and relation extraction) in a sequence, treating coreference resolution as an external black box. 
While word and span representations are shared across all tasks and updated to minimize multi-task loss, the model trains each task on gold input. Figure \ref{model_architecture_diagram} summarizes the baseline model's end-to-end architecture, and highlights the places where we propose improvements for our CitationIE model.

\paragraph{Feature Extraction}
The model extracts features from raw text in two stages. First, contextualized word embeddings are obtained for each section by running SciBERT \citep{Beltagy2019SciBERTAP} on that section of text (up to 512 tokens). Then, the embeddings from all words over all sections are passed through a bidirectional LSTM \cite{Graves2005BidirectionalLN} to contextualize each word's representation with those from other sections.

\paragraph{Mention Identification}
The baseline model treats this named entity recognition task as an IOBES sequence tagging problem \cite{Reimers2017OptimalHF}. The tagger takes the SciBERT-BiLSTM \citep{Beltagy2019SciBERTAP, Graves2005BidirectionalLN} word embeddings (as shown in the Figure \ref{model_architecture_diagram}), feeds them through two feedforward networks (not shown in Figure \ref{model_architecture_diagram}), and produces tag potentials at each word. These are then passed to a CRF \cite{Lafferty2001ConditionalRF} which predicts discrete tags.

\paragraph{Span Embeddings}
 For a given mention span, its span embedding is produced via additive attention \citep{Bahdanau2015NeuralMT} over the tokens in the span.

\paragraph{Coreference}
Using an external model, pairwise coreference predictions are made for all entity mentions, forming coreference clusters.

\paragraph{Salient Entity Classification}
Saliency is a property of entity clusters, but it is first predicted at the entity mention level. Each entity mention's span embedding is simply passed through two feedforward networks, giving a binary saliency prediction.

To turn these mention-level predictions into cluster-level predictions, the predicted saliency scores are max-pooled over all mentions in a coreference cluster to give cluster-level saliency scores.

\paragraph{Relation Extraction}
The model treats relation extraction as binary classification, taking as input a set of 4 typed salient entity clusters. For each entity cluster in the relation, per-section entity cluster representations are computed by taking the set of that entity's mentions in a given section, and max-pooling over the span embeddings of these mentions. The four entity-section embeddings (one for each entity in the relation) are then concatenated and passed through a feedforward network to produce a relation-section embedding. Then, the relation-section embeddings are averaged over all sections and passed through another feedforward network which returns a binary prediction.

\section{Citation-aware SciIE Dataset}

Although citation network information has been shown to be effective in other tasks, few works have recently tried using it in SciIE systems. One potential reason is the lack of a suitable dataset.

Thus, as a first contribution of this paper, we address this bottleneck by constructing a SciIE dataset that is annotated with citation graph information.\footnote{We have released code to construct this dataset: \url{https://github.com/viswavi/ScigraphIE}}
Specifically, we combine the rich annotations of SciREX with a source of citation graph information, S2ORC \citep{lo-etal-2020-s2orc}.
For each paper, S2ORC includes parsed metadata about which other papers cite this paper, which other papers are cited by this paper, and locations in the body text where reference markers are embedded.

To merge SciREX with S2ORC, we link records using metadata obtained via the Semantic Scholar API:\footnote{\url{https://www.semanticscholar.org/}} paper title, DOI string, arXiv ID, and Semantic Scholar Paper ID. For each document in SciREX, we check against all 81M documents in S2ORC for exact matches on any of these identifiers, yielding S2ORC entries for 433 out of 438 documents in SciREX. The final mapping is included in our repository for the community to use. Though our work only used the SciREX dataset, our methods can be readily extended to other SciIE datasets (including those mentioned in \S\ref{sec:task_definition}) using our released software.

\begin{figure}[t]
\centering
    \includegraphics[width=4.0cm]{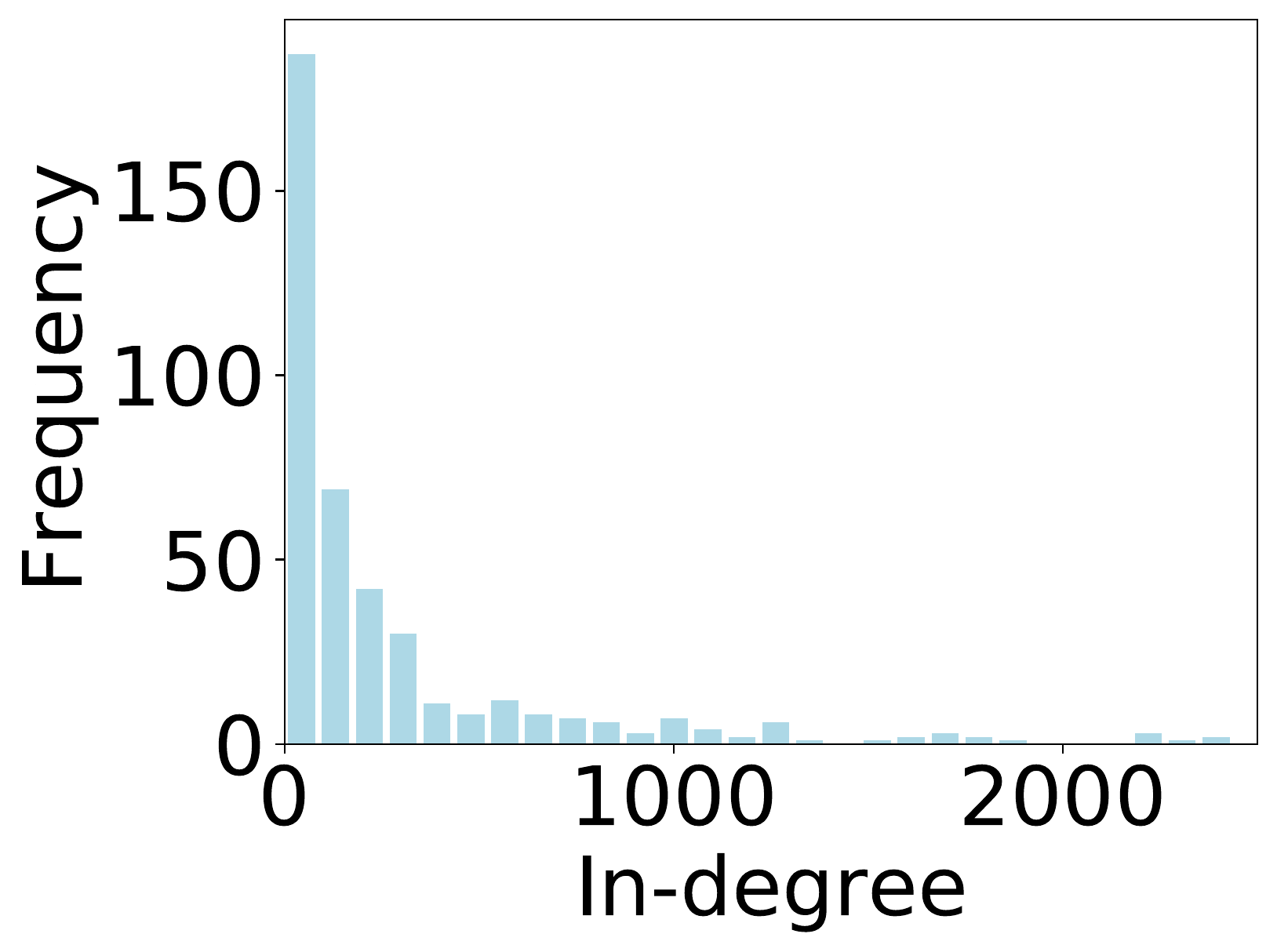}
    \includegraphics[width=3.6cm]{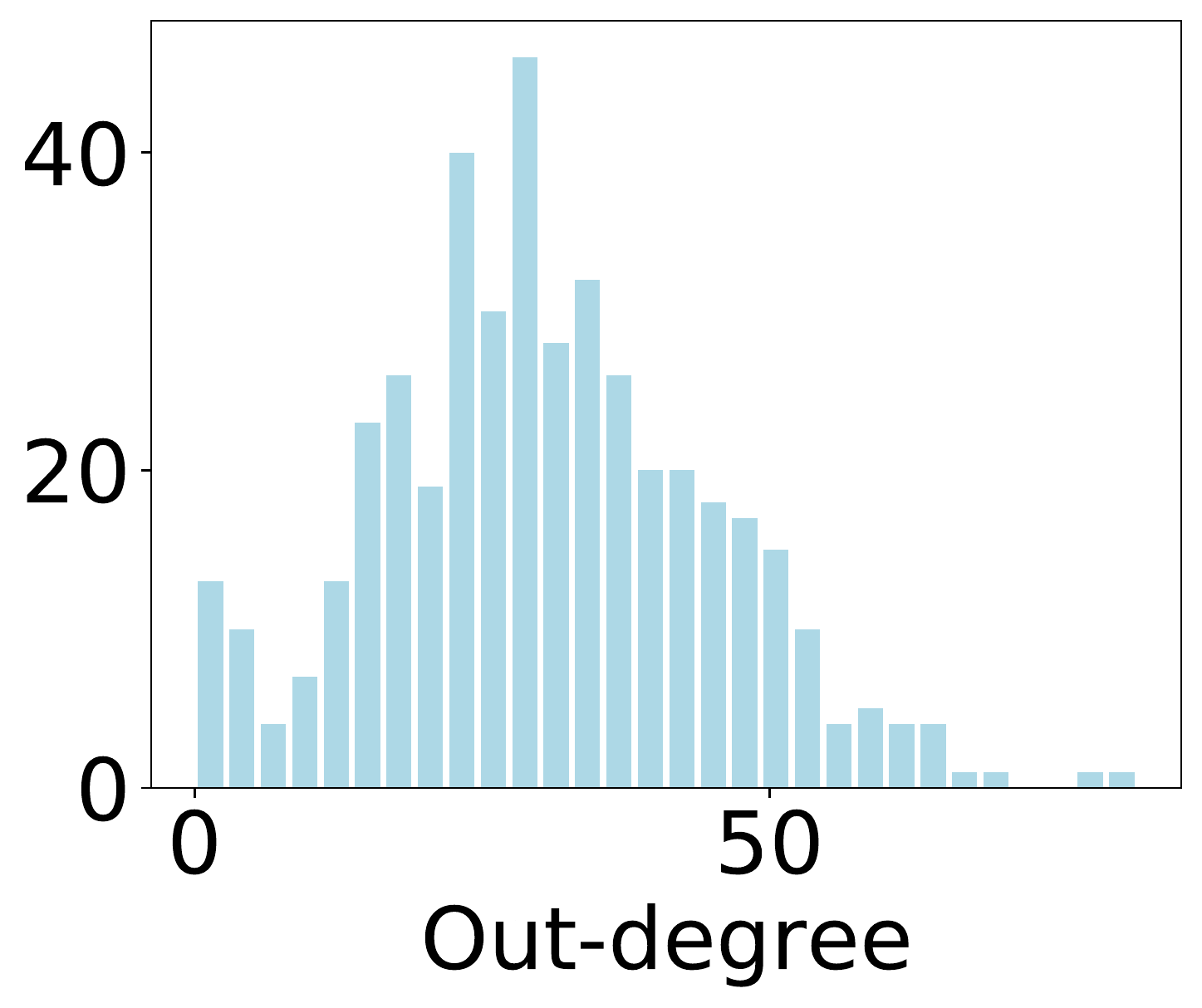}
    \caption{Degree statistics of SciREX documents in the citation graph.}
    \label{fig:degree_histogram}

\end{figure}

\paragraph{Statistics}
Examining the distribution of citations for all documents in the SciREX dataset (in Figure~\ref{fig:degree_histogram}), we observe a long-tailed distribution of citations per paper, and a bell-shaped distribution of references per paper.

In addition to the 5 documents we could not match to the S2ORC citation graph, 7 were incorrectly recorded as containing no references and 5 others were incorrectly recorded as having no citations. These errors are due to data issues in the S2ORC dataset, which relies on PDF parsers to extract information \cite{lo-etal-2020-s2orc}.

\section{CitationIE}
We now describe our citation-aware scientific IE architecture, which incorporates citation information into mention identification, salient entity classification, and relation extraction.
For each task, we consider two types of citation graph information, either separately or together: (1) \textit{structural information} from the graph network topology and (2) \textit{textual information} from the content of citing and cited documents.

\subsection{Structural Information}
\label{sec:structural_information}
The structure of the citation graph can contextualize a document within the greater body of work.

Prior works in scientific information extraction have predominantly used the citation graph only to analyze the content of citing papers, such as \emph{CiteTextRank} \citep{citetextrank} and \emph{Citation TF-IDF} \citep{caragea-etal-2014-citation}, which is described in detail in \S\ref{sec:citation_tf_idf}. However, the citation graph can be used to discover relationships between non-adjacent documents in the citation graph; prior works struggle to capture these relationships.

\citet{Arnold2009InformationEA} are the only prior work, to our knowledge, to explicitly use the citation graph's structure for scientific IE. They predict key entities related to a paper via random walks on a combined knowledge-and-citation-graph consisting of papers and entities, without considering a document's content. This approach is simple but cannot generalize to new or unseen entities.

 A rich direction of recent work has studied learned representations of networks, such as social networks \cite{perozzi2014deepwalk} and citation graphs \cite{cora_dataset, Yang2015NetworkRL, Bui2018NeuralGL, comparing_nrl}.
 In this paper, we show citation graph embeddings can improve scientific information extraction.
 


\paragraph{Construction of Citation Graph} 
To construct our citation graph, we found all nodes in the S2ORC citation graph within 2 undirected edges of any document in the SciREX dataset, including all edges between those documents. This process took 10 hours on one machine due to the massive size of the full S2ORC graph, resulting in a graph with $\sim$1.1M nodes and $\sim$5M edges.

\paragraph{Network Representation Learning}
We learn representations for each node (paper) using DeepWalk\footnote{An empirical comparison by \citet{comparing_nrl} found DeepWalk to be quite competitive on two citation graph node classification datasets, despite its speed and simplicity.} \citep{perozzi2014deepwalk} via the GraphVite library \citep{zhu2019graphvite}, resulting in a 128-dimensional ``graph embedding" for each document in our dataset. For each task, we incorporate the document-level graph embedding into that task's model component, by simply concatenating the document's graph embedding with the hidden state in that component. We do not update the graph embedding values during training.

\begin{figure}[t]
\centering
    \includegraphics[width=0.7\linewidth]{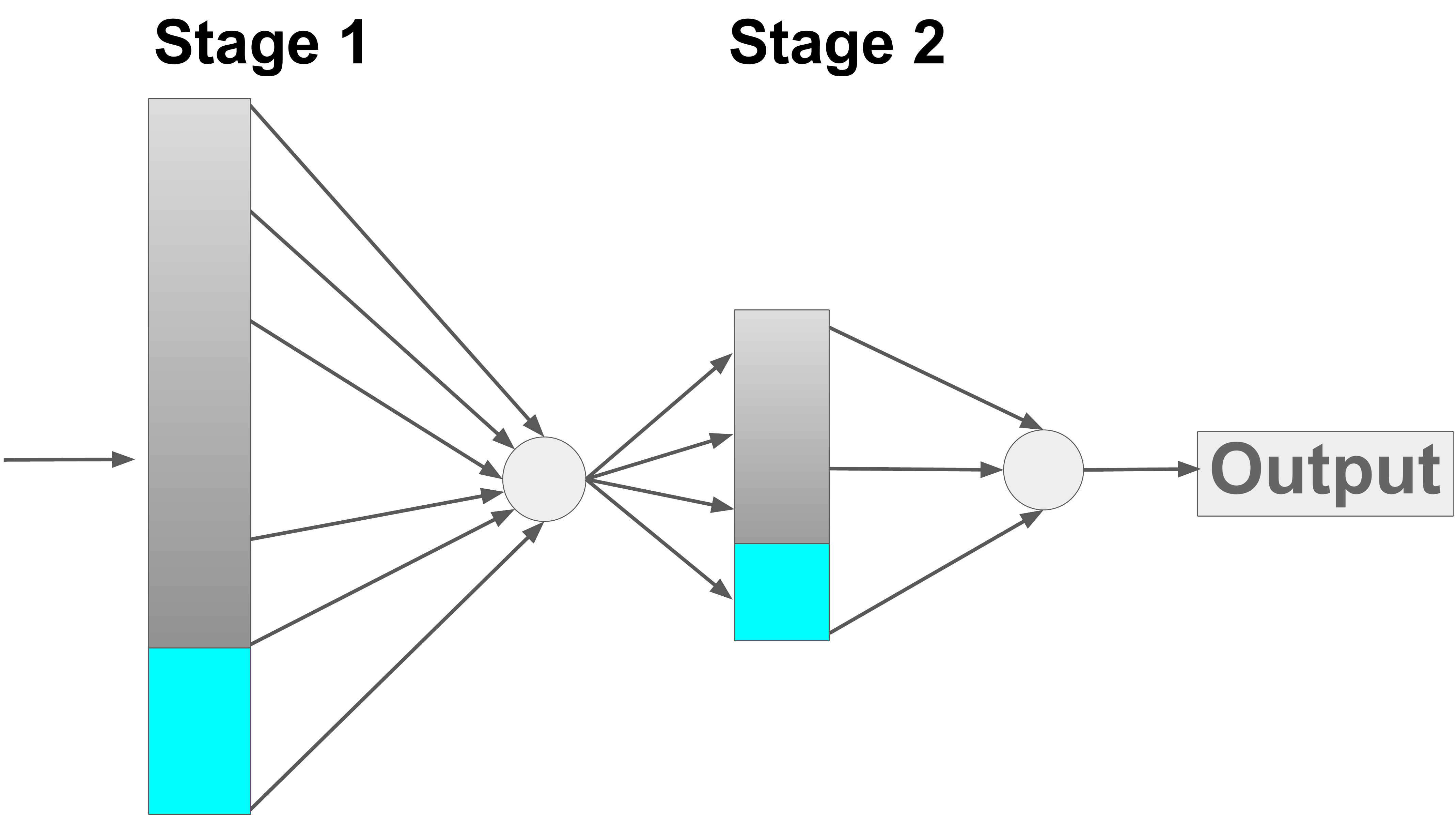}
    \caption{Feedforward architecture in each task (with CitationIE-specific parameters shown in light blue).}
    \label{fig:zoomed_in_model_architecture_graph}
\end{figure}

\paragraph{Incorporating Graph Embedding}
Each task in our CitationIE system culminates in a pair of feedforward networks. Figure \ref{fig:zoomed_in_model_architecture_graph} describes this general architecture,
though the input to these networks varies from task to task (SciBERT-BiLSTM embeddings for mention identification, span embeddings for salient entity classification, and per-section relation embeddings for relation extraction).

This architecture gives two options for where to concatenate the graph embedding into the hidden state - Stage 1 or Stage 2 - marked with a light blue block in Figure \ref{fig:zoomed_in_model_architecture_graph}. Intuitively, concatenating the graph embedding in a later stage feeds it more directly into the final prediction.
We find Stage 1 is superior for relation extraction, and both perform comparably for salient entity classification and mention identification. We give details on this experiment in Appendix \ref{sec:embedding_fusion}.

\subsection{Textual Information}
Most prior work using the citation graph for SciIE has focused on using the text of citing papers. We examine how to use two varieties of textual information related to citations.

\subsubsection{Citances}
Citation sentences, also known as ``citances" \citep{Nakov04citances}, provide an additional source of textual context about a paper. They have seen use in automatic summarization \citep{scisummnet}, but not in neural information extraction.

\begin{figure}[t]
\centering
    \includegraphics[width=0.7\linewidth]{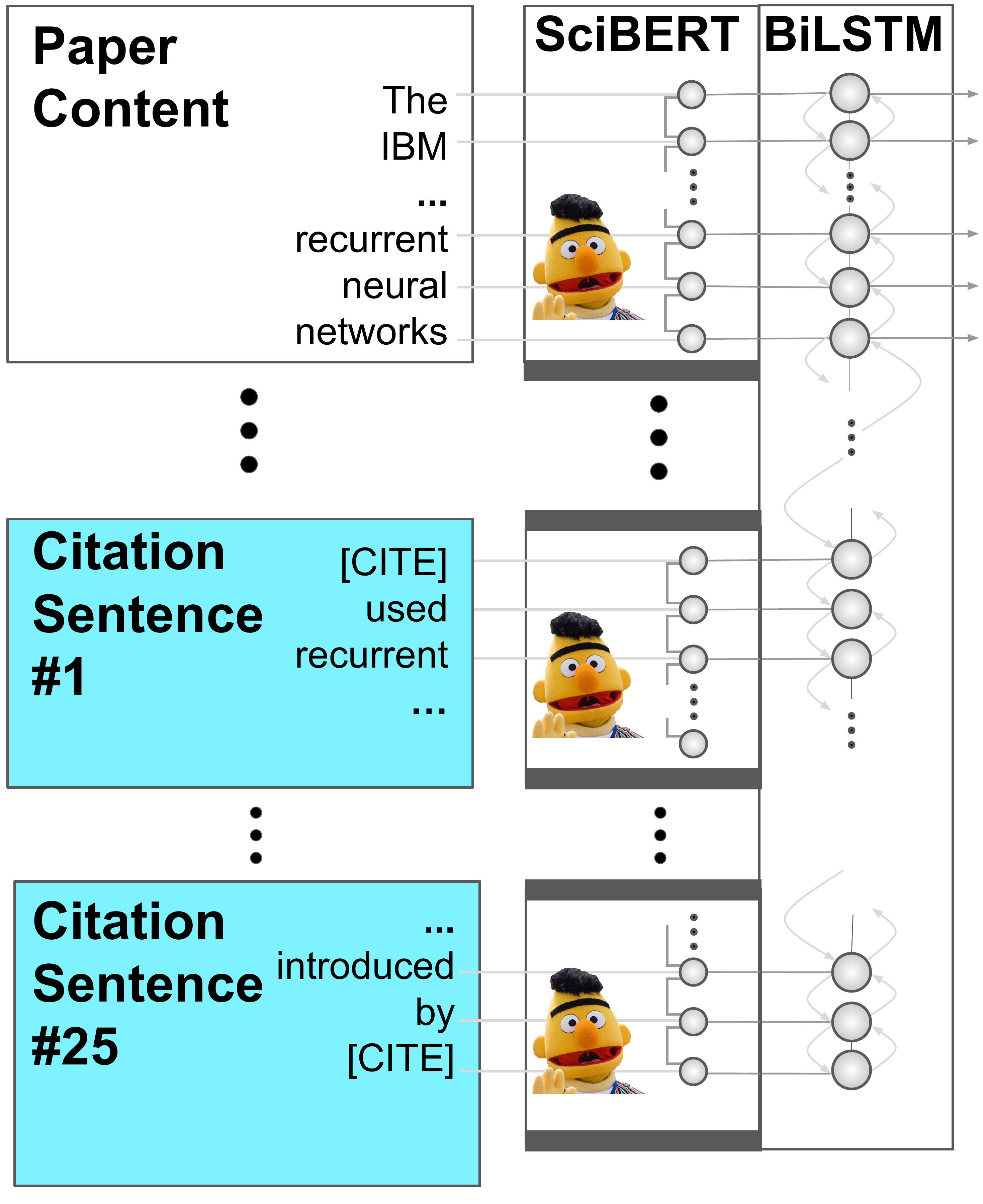}
    \caption{Incorporating citances into the text representation extractor.}
    \label{fig:zoomed_in_model_architecture_citances}
\end{figure}

In our work, we augment each document in our training set with its citances, treating each citance as a new section in the document. In this way, we incorporate citances into our CitationIE model through the shared text representations used by each task in our system, as shown in Figure \ref{fig:zoomed_in_model_architecture_citances}. If our document has many citations, we randomly sample 25 to use. For each citing document, we select citances centered on the sentence containing the first reference marker pointing to our document of interest, and include the subsequent and consequent sentences if they are both in the same section.

We ensure the mention identification step does not predict entities in citance sections, which would lead to false positive entities in downstream tasks. 

\subsubsection{Citation TF-IDF}
\label{sec:citation_tf_idf}
\emph{Citation TF-IDF} \citep{caragea-etal-2014-citation}, is a feature representing the TF-IDF value \citep{jones1972statistical} of a given token in its document's citances. We consider a variant of this feature: for each token in a document, we compute the TF-IDF of that token in each citance of the document, and average the per-citance TF-IDF values over all citances.
We implemented this feature only for saliency classification, as it explicitly reasons about the significance of a token in citing texts. As a local token-level feature, it also does not apply naturally to relation extraction, which operates on entire clusters of spans.

\subsection{Graph Structure and Text Content}
We lastly consider using graph embeddings and citances together in a single model for each task. We do this naively by including citances with the document's input text when first computing shared text features, and then concatenating graph embeddings into downstream task-specific components.

\section{Experiments}

\subsection{Metrics, Baselines and Training}

\subsubsection{Metrics}

The ultimate product of our work is an end-to-end document-level relation extraction system, but we also measure each component of our system in isolation, giving end-to-end and per-task metrics. All metrics, except where stated otherwise, are the same as described by \citet{jain-etal-2020-scirex}.

\paragraph{Mention Identification}
We evaluate mention identification with the average F1 score of classifying entities of each span type.

\paragraph{Salient Entity Classification}
Similar to \citet{jain-etal-2020-scirex} we evaluate this task at the mention level and cluster level. We evaluate both metrics on gold standard entity recognition inputs.

\paragraph{Relation Extraction}
This is the ultimate task in our pipeline. We use its output and metrics to evaluate the end-to-end system, but also evaluate relation extraction separately from upstream components to isolate its performance.
We specifically consider two types of metrics:

\noindent (1) \textit{Document-level}: For each document, given a set of ground truth 4-ary relations, we evaluate a set of predicted 4-ary relations as a sequence of binary predictions (where a matching relation is a true positive). We then compute precision, recall, and F1 scores for each document, and average each over all documents. We refer to this metric as the ``\textit{document-level}'' relation metric. To compare with \citet{jain-etal-2020-scirex}, this is the primary metric to measure the full system.

\noindent (2) \textit{Corpus-level}: When evaluating the relation extraction component in isolation, we are also able to use a more standard ``\textit{corpus-level}" binary classification evaluation, where each candidate relation from each document is treated as a separate sample. 

We also run both these metrics on a binary relation extraction setup, by flattening each set of 4-ary relations into a set of binary relations and evaluating these predictions as an intermediate metric.

\subsubsection{Baselines}
For each task, we compare against \citet{jain-etal-2020-scirex}, whose architecture our system is built on. No other model to our knowledge performs all the tasks we consider on full documents.
For the 4-ary relation extraction task, we also compare against the DocTAET model \citep{hou-etal-2019-identification}, which is considered as state-of-the-art for full-text scientific relation extraction \citep{jain-etal-2020-scirex, hou-etal-2019-identification}.

\paragraph{Significance}
To improve the rigor of our evaluation, we run significance tests for each of our proposed methods against its associated baseline, via paired bootstrap sampling \citep{koehn-2004-statistical}. In experiments where we trained multiple models with different seeds, we perform a hierarchical bootstrap procedure where we first sample a seed for each model and then sample a randomized test set.

\subsubsection{Training Details}
We build our proposed CitationIE methods on top of the SciREX repository\footnote{\url{https://github.com/allenai/SciREX}} \citep{jain-etal-2020-scirex} in the AllenNLP framework \citep{gardner-etal-2018-allennlp}.

For each task, we first train that component in isolation from the rest of the system to minimize the task-specific loss. We then take the best performing modifications and use them to train end-to-end IE models to minimize the sum of losses from all tasks. We train each model on a single GPU with batch size 4 for up to 20 epochs. We include detailed training configuration information in Appendix \ref{sec:training_configurations}.

For saliency classification and relation extraction, we trained the baseline and the strongest proposed models three times,\footnote{See Appendix \ref{sec:training_configurations} for exact seeds used} to improve reliability of our results. For mention identification, we did not retrain models, as the first set of results strongly suggested our proposed methods were not helpful.


\subsection{Quantitative Results}

\paragraph{Mention Identification}
For mention identification, we observe no major performance difference from using citation graphs, and include full results in Appendix \ref{sec:mention_identification_results}.

\paragraph{Salient Entity Classification}
\begin{table}[t] 
  \small
  \centering
  \begin{tabular}{lccc}
  \toprule 
  \textbf{Model} & F1 & P & R \\ \midrule
   \multicolumn{4}{c}{Salient Mention Evaluation} \\ \midrule
   Baseline (reported) & 57.9 & 57.5 & 58.4 \\
   Baseline (reimpl.) & 57.5 & 50.5 & 66.8 \\
   \hdashline 
   \emph{CitationIE} & & & \\
   w/ Citation-TF-IDF & 57.1 & 50.2 & 66.1 \\
   w/ Citances & 58.7$\dag$ & 51.4 & \textbf{68.5}$\dag$\\
   w/ Graph Embeddings & \textbf{59.2}$\dag$ & \textbf{53.5}$\dag$ & 66.3 \\
   w/ Graph + Citance & 58.4$\dag$ & 51.3 & 67.8$\dag$\\ \midrule
   \multicolumn{4}{c}{Salient Entity Cluster Evaluation} \\ \midrule
   Baseline (reimpl.) & 39.1 & 28.5 & \textbf{75.8} \\
   \hdashline 
   \emph{CitationIE} & & & \\
   w/ Citation-TF-IDF & 38.6 & 28.4 & 74.3 \\
   w/ Citances & 38.7 & 28.2 & 74.8 \\
   w/ Graph Embeddings & \textbf{40.3} & \textbf{29.8} & 74.5\\
  \bottomrule
  \end{tabular}
  \caption{Salient entity classification results.
  Baseline \citep{jain-etal-2020-scirex} and Graph Embedding model evaluations are each trained with 3 different model seeds, then metrics averaged; rest are from single model due to computational limitations. $\dag$ indicates significance at 95\% confidence. Best model is in bold for each metric.
  }
  \label{saliency-results-table} 
\end{table}

\setlength{\tabcolsep}{14pt}
\begin{table*}[t] 
\centering
\small
  \begin{tabular}{lccclccc}
  \toprule \textbf{Model} & F1 & P & R  & & F1 & P & R \\ \midrule
  & \multicolumn{7}{c}{4-ary Relation Extraction} \\ 
   & \multicolumn{3}{c}{Document-Level Metric} & & \multicolumn{3}{c}{Corpus-Level Metric}\\ \midrule
   Baseline (reported)\footnotemark & 57.0 & 82.0 & 44.0 & & N/A  & N/A  & N/A \\
   Baseline (reimpl.) & 49.8 & 50.1 & 50.1 & & 48.0 & 48.1 & 48.2 \\
    DocTAET & 65.5 & 62.4 & \textbf{85.1} & & 39.9 & 55.7 & 56.8 \\
    \hdashline
    \emph{CitationIE} \\
   w/ Citances & \textbf{69.2} & \textbf{70.0} & 76.6 & & 39.4 & 39.9 & 41.9 \\
   w/ Graph Embeddings & 68.5 & 67.5 & 76.2 & & \textbf{58.7}$\dag$ & \textbf{61.0}$\dag$ & \textbf{59.6}\\
   w/ Graph + Citance & 67.5 & 66.8 & 75.0 & & 51.9 & 54.6 & 54.5 \\ \midrule
  & \multicolumn{7}{c}{Binary Relation Extraction} \\ 
    & \multicolumn{3}{c}{Document-Level Metric} & & \multicolumn{3}{c}{ Corpus-Level Metric}\\ \midrule
   Baseline (reported) & 61.1 & 53.1 & 71.8 & & N/A  & N/A  & N/A \\
   Baseline (reimpl.) & 50.8 & 51.1 & 51.1 & & 41.2 & 48.4 & 44.6 \\
    \hdashline
    \emph{CitationIE} \\
   w/ Citances & 69.2 & 69.2 & \textbf{71.3} & & 43.3 & 46.7 & 44.0 \\
   w/ Graph Embeddings & \textbf{72.9} & \textbf{70.4} & 56.1 & & \textbf{51.0}$\dag$ & \textbf{54.1}$\dag$ & \textbf{57.1} \\
   w/ Graph + Citance & 66.2 & 65.9 & 68.1 & & 48.0$\dag$ &  51.4 & 52.7 \\ \bottomrule
  \end{tabular}
  \caption{Comparing methods on relation extraction. Baseline, Graph Embedding, and Graph + Citance models were evaluated over 3 model seeds, and the remainder with a single seed. We use Macro-F1 for corpus-level evaluation. $\dag$ indicates significance at 95\% confidence, and best implemented model in each metric is bolded. Graph embeddings significantly improve over baseline on 4-ary and binary corpus-level F1 ($p < 0.05$), but are less significant on document-level F1 metrics ($p \approx 0.11$).}
  \label{relation-only-table} 
\end{table*}
\footnotetext{ Reported as ``Component-wise Binary and 4-ary Relations" in \citet{jain-etal-2020-scirex}}

Table \ref{saliency-results-table} shows the results of our CitationIE methods. We observe:\\
(1) Using citation graph embeddings significantly improves the system with respect to the salient mention metric.\\
(2) Graph embeddings do not improve cluster evaluation significantly (at 95\%) due to the small test size\footnote{The limited size of this test set is an area of concern when using the SciREX dataset, and improving statistical power in SciIE evaluation is a crucial area for future work.} (66 samples) and inter-model variation. \\
(3) Incorporating graph embeddings and citances simultaneously is no better than using either. \\
(4) Our reimplemented baseline differs from the results reported by \citet{jain-etal-2020-scirex} despite using their published code to train their model. This may be because we use a batch size of 4 (due to compute limits) while they reported  a batch size of 50.

\paragraph{Relation Extraction}
Table \ref{relation-only-table} shows that using graph embeddings here gives an 11.5 point improvement in document-level F1 over the reported baseline,\footnote{The large gap between reimplemented and reported baselines is likely due to our reproduced results averaging over 3 random seeds. When using the same seed used by \citet{jain-etal-2020-scirex}, the baseline's document-level test F1 score is almost 20 points better than with two other random seeds.
} and statistically significant gains on both corpus-level F1 metrics.

Despite seemingly large gains on the document-level F1 metric, these are not statistically significant due to significant inter-model variability and small test set size, despite the graph embedding model performing best at every seed we tried.

\paragraph{End-to-End Model}
\setlength{\tabcolsep}{10pt}
\begin{table}[t] 
\centering
\small
  \begin{tabular}{llll}
  \toprule \textbf{Model} & F1 & P & R \\ \midrule
   & \multicolumn{3}{c}{4-ary Relation Extraction}\\ \midrule
   Baseline (reported) & 0.8 & 0.7 & 17.3 \\
   Baseline (reimpl.) & 0.44 & 0.23 & \textbf{22.66} \\
    \hdashline
    \emph{CitationIE} \\
   w/ Graph Embeddings & \textbf{1.48} & 1.31 & 20.04\\
   w/ Citances & 0.75 & \textbf{7.03} & 13.36\\ \midrule
    & \multicolumn{3}{c}{Binary Relation Extraction} \\ \midrule
   Baseline (reported) & 9.6 & 6.5 & 41.1 \\
   Baseline (reimpl.) & 6.48 & 4.09 & 43.83 \\
    \hdashline
    \emph{CitationIE} \\
   w/ Graph Embeddings & \textbf{7.70} & \textbf{5.42} & 37.17 \\ 
   w/ Citances & 7.61 & 4.97 & \textbf{43.57} \\\bottomrule
  \end{tabular}
  \caption{End-to-end model evaluation. Each model was evaluated over 3 model seeds.}
  \label{end-to-end-table} 
\end{table}
From Table \ref{end-to-end-table}, we observe:\\
(1) Using graph embeddings appears to have a positive effect on the main task of 4-ary relation extraction. However, these gains are not statistically significant ($p=0.235$) despite our proposed method outperforming the baseline at every seed, for the same reasons as mentioned above. \\
(2) On binary relation evaluation, we observe smaller improvements which had a lower p-value ($p=0.099$) due to lower inter-model variation. \\
(3)  Using citances instead of graph embeddings still appears to outperform the baseline (though by a smaller margin than the graph embeddings).
\subsection{Analysis}
\label{Sec:analysis}
\begin{figure}[t]
    \includegraphics[width=4.1cm]{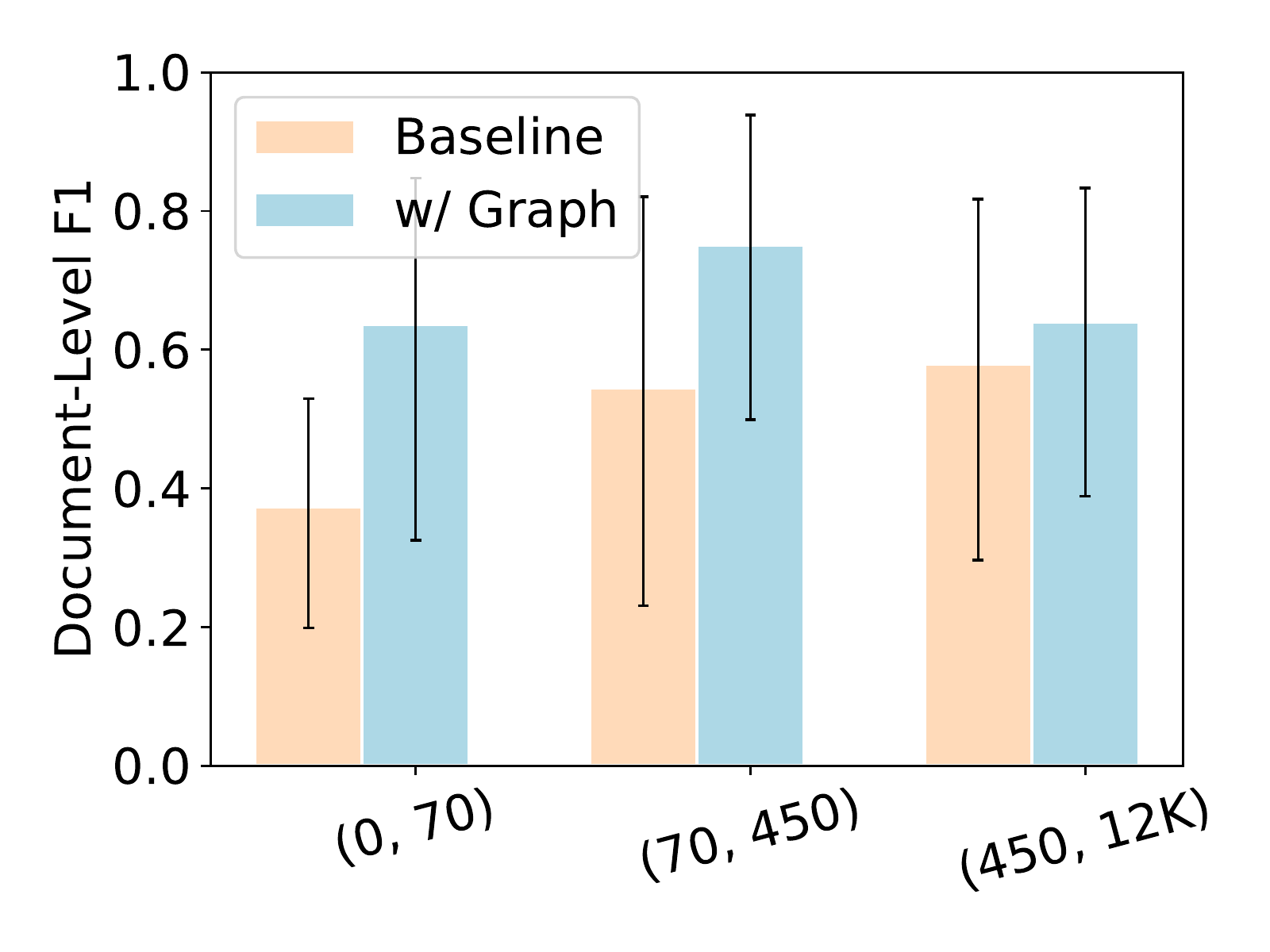}
   \hspace{-0.55cm} \includegraphics[width=4.0cm]{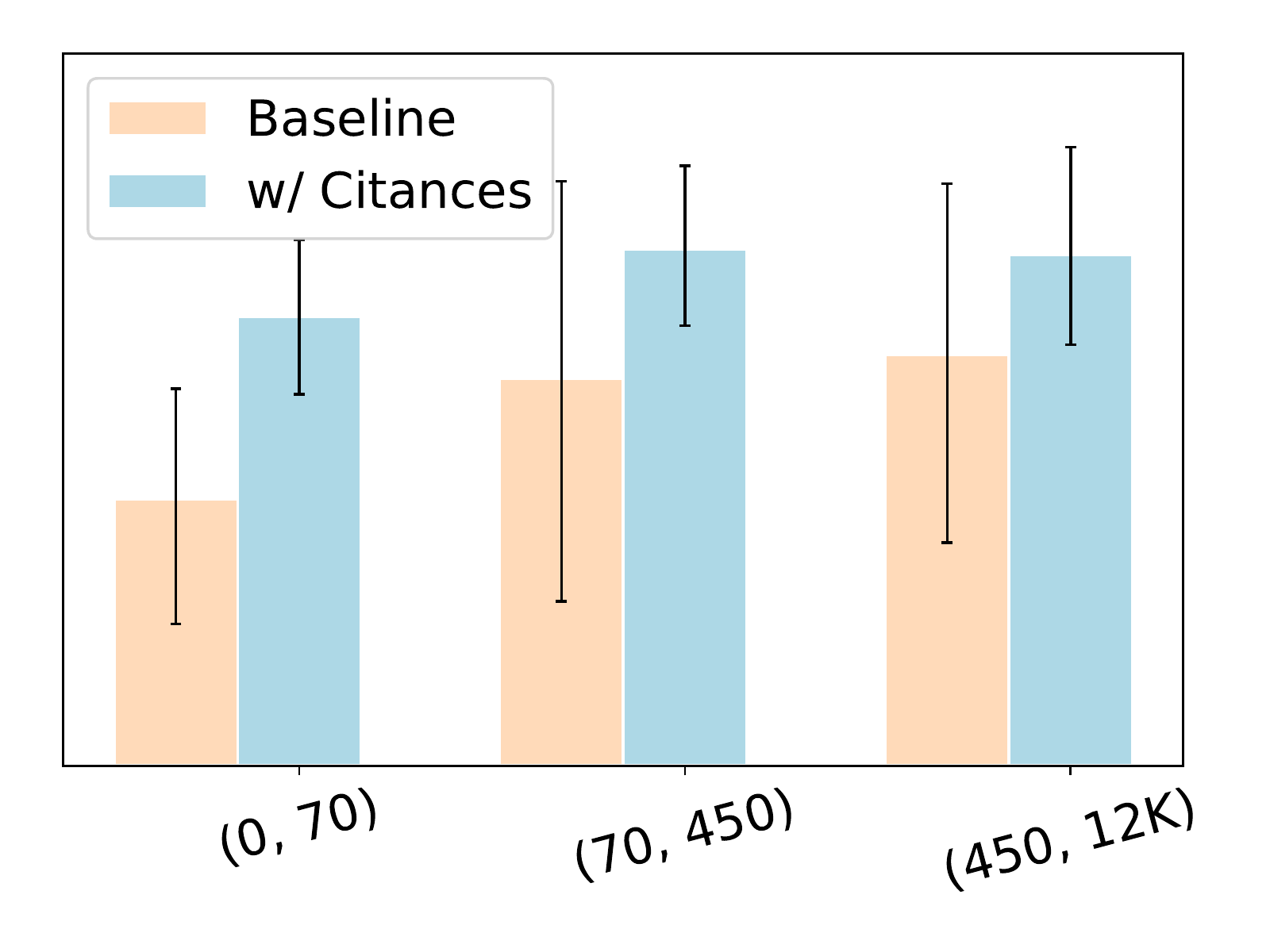}
    \caption{Document-level relation extraction F1 score of CitationIE models with graph embeddings (left) and citances (right), compared with the baseline (red) on documents grouped by number of citations. 
    }    \label{bucketed_eval_on_citation_graph_degree}
\end{figure}

We analyzed our experimental results, guided by the following four questions:

\paragraph{Do papers with few citations benefit from citation graph information?}
Our test set only contains two documents with zero citations, so we cannot characterize performance on such documents. However, Figure \ref{bucketed_eval_on_citation_graph_degree} shows that the gains provided by the proposed CitationIE model with graph embeddings counterintuitively shrink as the number of citations of a paper increases. We also observe this with citances, to a lesser extent. This suggests more work needs to be done to represent citation graph nodes with many edges.

\begin{figure}[t]
\centering
    \hspace{-0.3cm}\includegraphics[width=4.2cm]{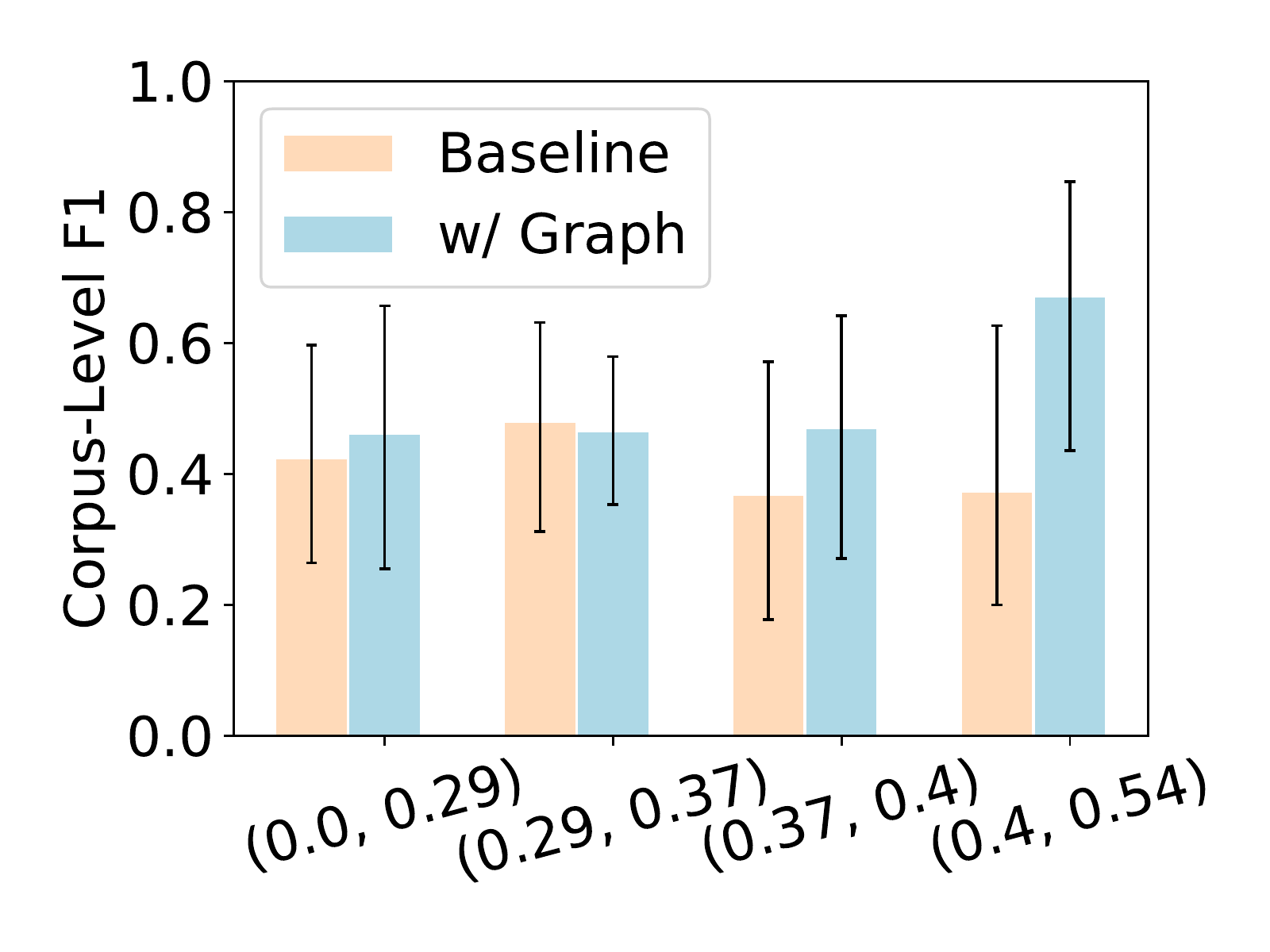}
    \hspace{-0.6cm}\includegraphics[width=4.1cm]{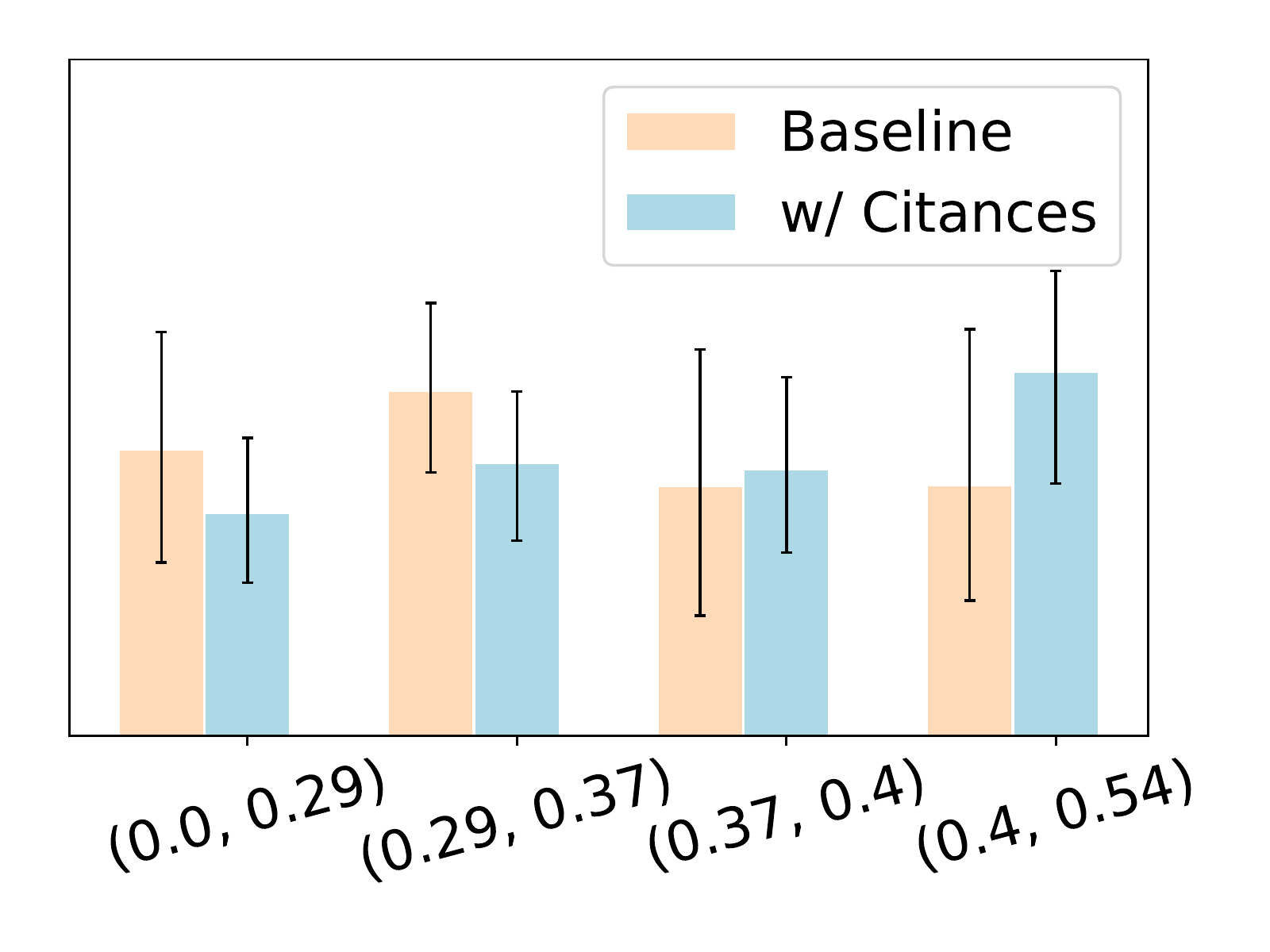}\hspace{0.2cm}
    \caption{Corpus-Level F1 of relation extraction models, bucketed by the average distance between entity mentions in each relation.}    \label{bucketed_relation_eval_on_cluster_distance}
\end{figure}

\paragraph{How does citation graph information help relation extraction?}
With relation extraction, we found citation graph information provides strongest gains when classifying relations between distant entities in a document, seen in Figure \ref{bucketed_relation_eval_on_cluster_distance}. For each relation in the test set, we computed the average distance between pairs of entity mentions in that relation, normalized by total document length. We  find models with graph embeddings or citances perform markedly better when these relations span large swaths of text. This is particularly useful since neural models still struggle to model long-range dependencies effectively \citep{brown2020language}.

\paragraph{Does citation graph information help contextualize important terms?}
Going back to our motivating example of a speech paper referring to ImageNet in passing \S\ref{case_study_example}, we hypothesized that adding context from citations helps deal with terms that are important in general, but not for a given document.

To measure this, we grouped all entities in our test dataset by their ``global saliency rate" measured on the test set: given a span, what is the probability that this span is salient in any given occurrence? 
\begin{figure}[t]
    \centering
    \hspace{-0.3cm}\includegraphics[width=4.3cm]{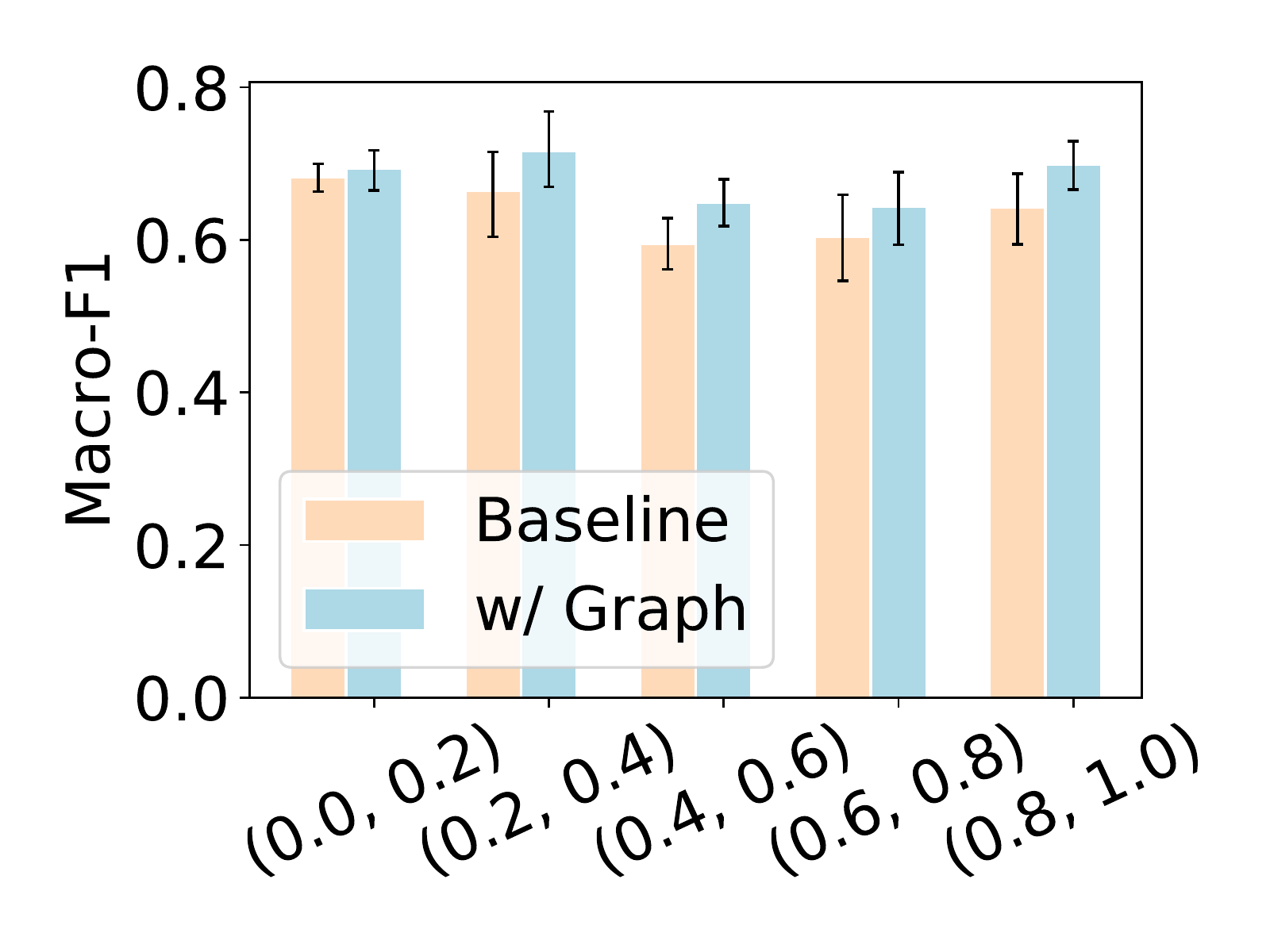}
    \hspace{-0.6cm}\includegraphics[width=4.2cm]{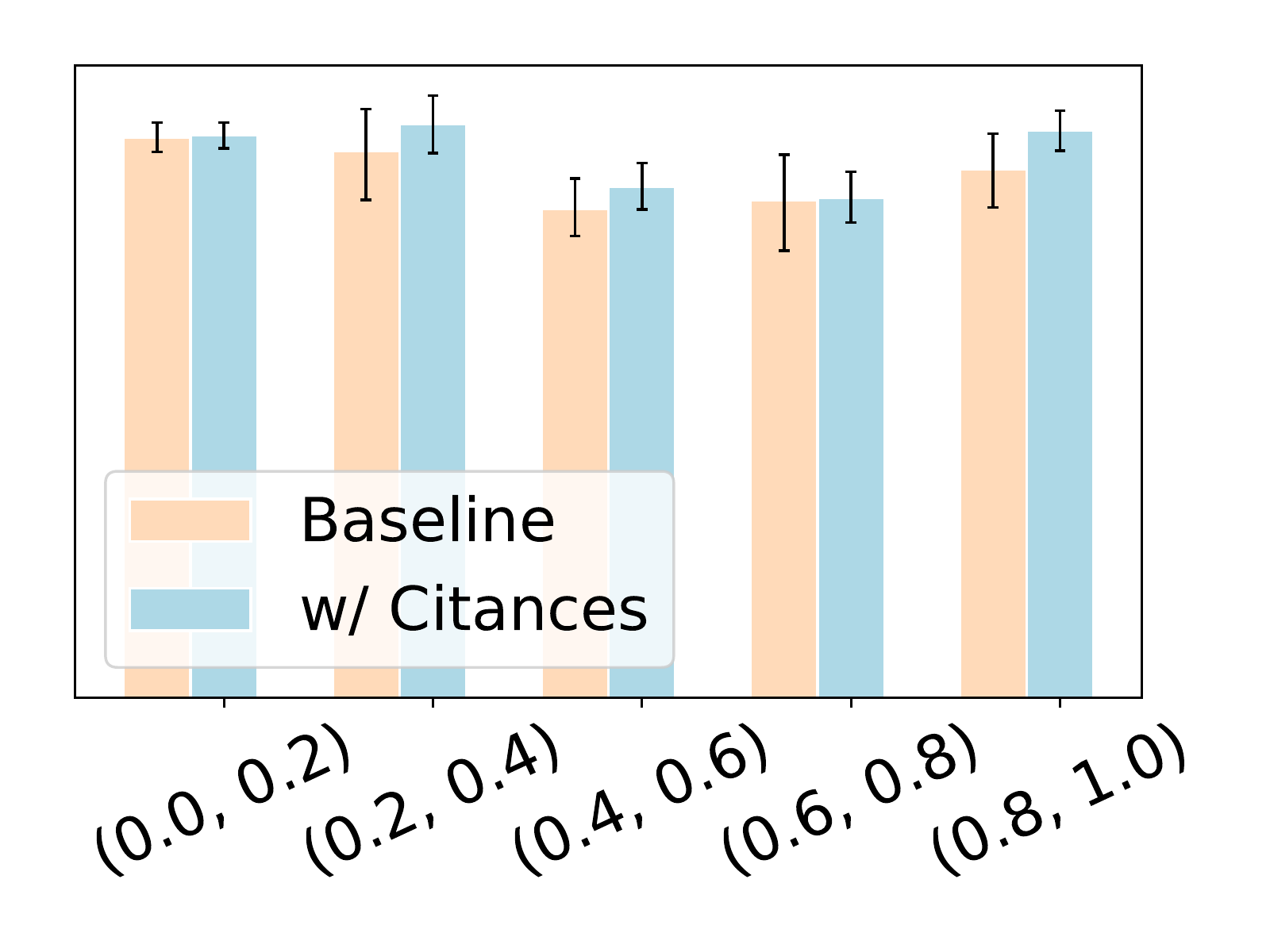}
    \caption{Macro F1 of salient mention classification models, evaluated on test-set spans, each bucketed by their training-set global saliency rate.}
    \label{bucketed_salient_eval_on_global_saliency_rate}
\end{figure}

In Figure \ref{bucketed_salient_eval_on_global_saliency_rate}, we observe that most of the improvement from graph embeddings and citances comes at terms which are labeled as salient in at least 20\% of their training-set mentions. This suggests that citation graph information yields improvements with reasoning about important terms, without negatively interfering with less-important terms.

\section{Implications and Future Directions}
We explore the use of citation graph information in neural scientific information extraction with \emph{CitationIE}, a model that can leverage either the structure of the citation graph or the content of citing or cited documents. We find that this information, combined with document text, leads to particularly strong improvements for salient entity classification and relation extraction, and provides an increase in end-to-end IE system performance over a strong baseline.

Our proposed methods reflect some of the simplest ways of incorporating citation graph information into a neural SciIE system.
As such, these results can be considered a proof of concept.
In the future we will explore ways to extract richer information from the graph using more sophisticated techniques, hopefully better capturing the interplay between citation graph structure and content.
Finally, we evaluated our proof of concept here on a single dataset in the machine learning domain. While our methods are not domain-specific, verifying that these methods generalize to other scientific domains is important future work.

\section*{Acknowledgments}

The authors thank Sarthak Jain for assisting with reproducing baseline results, Bharadwaj Ramachandran for giving advice on figures, and Siddhant Arora and Rishabh Joshi for providing suggestions on the paper.
The authors also thank the anonymous reviewers for their helpful comments. 
This work was supported by the Air Force Research Laboratory under agreement number FA8750-19-2-0200. The U.S. Government is authorized to reproduce and distribute reprints for Governmental purposes notwithstanding any copyright notation thereon. The views and
conclusions contained herein are those of the authors and should not be interpreted as necessarily representing the official policies or endorsements, either expressed or implied, of the Air Force Research Laboratory or the U.S. Government.




\bibliography{acl2020}
\bibliographystyle{acl_natbib}

\appendix
\section{Appendices}
\subsection{Training Configurations}
\label{sec:training_configurations}
We train each model on a single 11GB NVIDIA GeForce RTX 2080 Ti GPU with a batch size of 4. We train for up to 20 epochs, and set the \texttt{patience} parameter in AllenNLP to 10; if the validation metric does not improve for 10 consecutive epochs, we stop training early. For each task-specific model, we use a product of validation loss and corpus-level binary F1 score on the validation set as the validation metric. For salient entity classification and relation extraction, we choose the best threshold on the validation set using F1 score.

In total, training with these configurations takes roughly 2 hours for salient entity classification, 8 hours for mention identification, 18-24 hours for relation extraction, and 24-30 hours for the end-to-end system. Our CitationIE models took roughly as long to train as the baseline SciREX models did.

For models that we trained three different times, we use different seeds for each software library:
\begin{itemize}
    \item For PyTorch, we use seeds 133,\footnote{133/1337/13370 is the default seed setting in AllenNLP.} 11, and 22
    \item For Numpy, we use seeds 1337, 111, and 222
    \item For Python's \texttt{random} library, we use seeds 11370, 1111, and 2222
\end{itemize}

\subsection{Mention Identification Results}
\label{sec:mention_identification_results}
\setlength{\tabcolsep}{6pt}

\begin{table}[h] 
  \centering
  \small
  \begin{tabular}{lccc}
  \toprule
   \textbf{Model} & F1 & P & R \\
   \midrule
  & \multicolumn{3}{c}{Mention Identification} \\ 
  \midrule
  Baseline (reported)\footnote{ We used the same evaluation code as the authors, so it is unclear why our evaluation indicate better results} & 70.7 & 71.7 & 71.2 \\
  \midrule
  Baseline (reimpl.) & \textbf{74.6}$\dag$ & 73.7 & \textbf{75.6}$\dag$ \\
  w/ Citances & 74.0 & 73.0 & 75.0 \\
  w/ Graph Embeddings  & 74.4 & \textbf{74.4}$\dag$ & 74.3 \\
  w/ Graph + Citance & 73.6 & 73.0 & 74.3 \\ 
  \bottomrule
  \end{tabular}
  \caption{Mention Identification Results. $\dag$ indicates significance at 95\% confidence. Best model is in bold for each metric.
  \label{ner-only-table} 
  }
\end{table}
We include results from using citation graph information for the mention identification task in Table \ref{ner-only-table}. We observe no major improvements in this task. Intuitively, recognizing a named entity in a document may not require global context about the document (e.g. ``LSTM" almost always refers to a \texttt{Method}, regardless of the paper where it is used), so the lack of gains in this task is unsurprising.

\subsection{Combining Graph Embeddings with Word Embeddings}
\label{sec:embedding_fusion}
Each of our task-specific components in the CitationIE model contains two feedforward networks where we may concatenate graph embedding information. We refer to these two options for where to fuse graph embedding information as "early fusion" and "late fusion", illustrated in Figure \ref{fig:zoomed_in_model_architecture_graph}.

Here we show a detailed comparison of early fusion vs late fusion models on Mention Identification (Table \ref{sec:fusion-comparison-ner}), Salient Entity Classification (Table \ref{sec:saliency_classification_fusion_results}), and Relation Extraction (Table \ref{sec:relation-only-fusion-table}). Based on these results, we used early fusion in our final CitationIE models for mention identification and relation extraction. For saliency classification, the relative performance of early fusion and late fusion differed across our two metrics, making this inconclusive. We used early fusion for saliency classification in the end-to-end model due to strong empirical performance there.

\label{sec:mention_identification_fusion_results}
\setlength{\tabcolsep}{6pt}

\begin{table}[h] 
  \centering
  \small
  \begin{tabular}{lccc}
  \toprule
   \textbf{Model} & F1 & P & R \\
   \midrule
  & \multicolumn{3}{c}{Mention Identification} \\ 
  \midrule
  Graph Embed. (early fusion) & \textbf{74.4}$\dag$ & \textbf{74.4}$\dag$ & 74.3 \\
  Graph Embed. (late fusion) & 74.1 & 73.1 & \textbf{75.1}$\dag$ \\ 
  \bottomrule
  \end{tabular}
  \caption{Comparing CitationIE models for mention identification with early graph embedding fusion vs late fusion. Results are shown from single-model evaluation. $\dag$ indicates significance at 95\% confidence. Best model is in bold for each metric.
  }
  \label{sec:fusion-comparison-ner} 
\end{table}

\begin{table}[h] 
  \small
  \centering
  \begin{tabular}{lccc}
  \toprule 
  \textbf{Model} & F1 & P & R \\ \midrule
   \multicolumn{4}{c}{Salient Mention Evaluation} \\ \midrule
   Graph Embed. (early fusion) & 57.1 & \textbf{54.4}$\dag$ & 60.1 \\
   Graph Embed. (late fusion) & \textbf{59.2}$\dag$ & 53.5 & \textbf{66.3}$\dag$\\ \midrule
   \multicolumn{4}{c}{Salient Entity Cluster Evaluation} \\ \midrule
   Graph Embed. (early fusion) & \textbf{43.3}$\dag$ & \textbf{33.8}$\dag$ & 72.0  \\
   Graph Embed. (late fusion) & 40.3 & 29.8 & \textbf{74.5}$\dag$ \\
  \bottomrule
  \end{tabular}
  \caption{Comparing CitationIE models for salient entity classification with early graph embedding fusion vs late fusion. The early fusion model was trained once, while late fusion numbers are reported over an average of 3 runs. $\dag$ indicates significance at 95\% confidence. Best model is in bold for each metric.
  }
  \label{sec:saliency_classification_fusion_results} 
\end{table}

\setlength{\tabcolsep}{14pt}
\begin{table*}[t] 
\centering
\small
  \begin{tabular}{lccclccc}
  \toprule \textbf{Model} & F1 & P & R  & & F1 & P & R \\ \midrule
  & \multicolumn{7}{c}{4-ary Relation Extraction} \\ 
   & \multicolumn{3}{c}{Document-Level Metrics} & & \multicolumn{3}{c}{Corpus-Level Metrics}\\ \midrule
   Graph Embeddings (early fusion) & \textbf{68.5} & \textbf{67.5} & \textbf{76.2} & & 58.7 & 61.0 & 59.6\\
   Graph Embeddings (late fusion) & 63.3 & 61.8 & 67.3 & & \textbf{75.8}$\dag$ & \textbf{76.0}$\dag$ & \textbf{76.1}$\dag$ \\
  & \multicolumn{7}{c}{Binary Relation Extraction} \\ 
    & \multicolumn{3}{c}{Document-Level Metrics} & & \multicolumn{3}{c}{ Corpus-Level Metrics}\\ \midrule
   Graph Embeddings (early Fusion) & \textbf{72.9} & \textbf{70.4} & 56.1 & & 51.0 & 54.1 & 57.1 \\
   Graph Embeddings (late fusion) & 58.3 & 58.0 & \textbf{59.0} & & \textbf{53.6} & \textbf{58.1}$\dag$ & \textbf{66.4} \\ \bottomrule
  \end{tabular}
  \caption{Comparing CitationIE models for relation extraction with early graph embedding fusion vs late fusion. Early fusion models were trained 3 times, late fusion was trained once. $\dag$ indicates significance at 95\% confidence, and the best model in each metric is bolded.}
  \label{sec:relation-only-fusion-table} 
\end{table*}

\end{document}